\documentclass{article}
\usepackage{url}
\usepackage{graphicx}
\usepackage{amssymb,amsmath}
\usepackage[stable]{footmisc}

\newcommand{\edonkey}{eDonkey}
\newcommand{\eg}{{\em e.g.}}

\newcommand{\gnutella}{{\em Gnutella}}
\newcommand{\ie}{{\em i.e.}}

\newcommand{\moins}{^-}
\newcommand{\moinsmoins}{^{-\,-}}

\newcommand{\plus}{^+}
\newcommand{\plusplus}{^{++}}
\newcommand{\primemoins}{'^{-}}
\newcommand{\primeplus}{'^{+}}

\newcommand{\tagged}{T}
\newsavebox{\fmbox}

\graphicspath{{Figures/}{Plots/}}

\begin{document}

\title{Quantifying Paedophile Activity\\ in a Large P2P System\footnote{A shorter version of this work has been published in \cite{QuantifyingInfocom}.}}

\author{
Matthieu Latapy, Cl\'emence Magnien, and Rapha\"el Fournier\\
CNRS and UPMC\\
4, place Jussieu 75005 Paris, France\\
\{{\em firstname.lastname}\}@lip6.fr
%\IEEEauthorblockN{Matthieu Latapy, Cl\'emence Magnien, and Rapha\"el Fournier}
%\IEEEauthorblockA{
%CNRS and UPMC\\
%4, place Jussieu 75005 Paris, France\\
%\{{\em firstname.lastname}\}@lip6.fr
%}
}

\date{\today}

%\markboth{Journal on Selected Areas in Communications}{}
\maketitle

\begin{abstract}
Increasing knowledge of paedophile activity in P2P systems is a crucial societal concern, with important consequences on child protection, policy making, and internet regulation. Because of a lack of traces of P2P exchanges and rigorous analysis methodology, however, current knowledge of this activity remains very limited. We consider here a widely used P2P system, eDonkey, and focus on two key statistics: the fraction of paedophile queries entered in the system and the fraction of users who entered such queries. We collect hundreds of millions of keyword-based queries; we design a paedophile query detection tool for which we establish false positive and false negative rates using assessment by experts; with this tool and these rates, we then estimate the fraction of paedophile queries in our data; finally, we design and apply methods for quantifying users who entered such queries. We conclude that approximately 0.25\,\% of queries are paedophile, and that more than 0.2\,\% of users enter such queries. These statistics are by far the most precise and reliable ever obtained in this domain.
\end{abstract}

%\begin{IEEEkeywords}
%Computer Networks, Measurement, Paedophile activity, \edonkey, Forensics.
%\end{IEEEkeywords}
%
%\IEEEpeerreviewmaketitle

\section{Introduction}
\label{sec-introduction}

%\IEEEPARstart{I}{t} 
It is widely acknowledged that peer-to-peer (P2P) file exchange systems host large amounts of paedophile content (mainly movies and pictures), which is a crucial societal concern. In addition to children victimisation, the wide availability of paedophile material is a great danger for regular users (including children and teenagers), who may be exposed unintentionally to extremely harmful content. In particular, this may lead initially innocent users to develop an interest in child pornography. It also has a strong impact on the public acceptance of paedophilia and induces a trivialisation of such content. Much work is devoted to these psychological and societal issues, see \cite{quayle08child,wolak06online}.

Downloading and/or providing paedophile content is a legal offence in many countries, and there is a correlation between downloading paedophile content and having actual sexual intercourse with children \cite{kim05from}. This makes fighting these exchanges a key issue for law enforcement \cite{wolak03internet,wortley06child}.  This also has much impact on P2P and internet regulation, and is used as a key allegation against people providing P2P facilities. For instance, people providing indexes of files available in P2P systems (including a small fraction of files with paedophile content) are often accused of helping and promoting paedophile exchanges, with strong penal threats \cite{delahunty06ed2k,murray05peer}.

For these reasons, knowledge of paedophile activity in P2P systems is a critical resource for law enforcement, child protection and policy making. See \cite{quayle08child,wolak03internet,wolak06online,wortley06child} for surveys on these issues.
However, current knowledge on this activity and its extent remains very limited and is subject to controversy \cite{hughes06is,koontz03file,steel09child,waters07child,wolak03internet,wolak06online}.

\medskip

In this paper, we provide ground truth on paedophile activity in a large P2P system,  at an unprecedented level of accuracy and reliability.  We focus on two basic yet crucial statistics: the fraction of paedophile queries entered in the system and the fraction of users entering such queries.  We establish reference methodology and tools for obtaining these values, and provide them in the case of the \edonkey\ system, which is one of the largest P2P systems currently used~\cite{schulze09ipoque}.

\medskip

Obtaining precise such information on paedophile activity in P2P systems
 raises several challenges:

\begin{itemize}

\item {\em Appropriate data collection.} Obtaining large-scale data of activity in P2P systems is a difficult task in itself. The main reasons are the lack of central authority, the size of these systems and their high dynamics, the poor structure of the traffic, and limited user identification.

\item {\em Paedophile activity identification.} As the relative am\-ount of paedophile activity in P2P systems is very low, quantifying it by manually inspecting a random sample of the data is not feasible: this sample would have to be very large in order to contain a significant amount of paedophile activity.  Moreover, this activity is often hidden (paedophiles use very specific keywords), and recognising it requires a deep expertise of the domain.  Finally, machine learning approaches, though appealing, cannot be applied in this context because of the lack of prior knowledge of representative paedophile data.

\item {\em Rigorous inference of statistics.} In a context where detection of paedophile activity as well as user identification are prone to errors, inferring reliable statistics is difficult. In addition, these statistics may fluctuate greatly with time, which makes their relevance unsure. Direct computations are not satisfactory to this regard, and the statistics must be carefully examined before concluding.

\end{itemize}

\medskip
To address these challenges, we make the following contributions:
\begin{itemize}
\item {\em Datasets.} We collect and publicly provide two sets of keyword-based queries entered by \edonkey\ users, on two different servers in 2007 and 2009. Each spans several weeks of activity (10 and 28, respectively) and contains hundreds of millions keyword-based queries, involving millions of users.  Using two datasets collected on different servers and at different dates increases the generality of our results significantly.
\item {\em Detection tool.} Using domain knowledge of paedophile keywords, we design and publicly provide a tool for automatic detection of paedophile queries. We evaluate its success rate by a rigorous assessment involving 21 experts having a deep knowledge of online paedophile activity. These experts work in various national and international law-enforcement agencies and well-established NGOs, including {\em Europol} and the {\em National Center for Missing \& Exploited Children}.
\item {\em Quantification.} Our tool detects hundreds of thousands paedophile queries in our datasets.  Using the error rates of the tool, we derive a reliable estimate of the actual fraction of paedophile queries they contain, which is approximately $0.25\,\%$.  We then design several complementary approaches to estimate the fraction of observed users who enter paedophile queries and check both their statistical significance and their consistence. We finally establish a lower bound of $0.2\,\%$ for users who enter paedophile queries in the 2007 dataset. Analysis of the 2009 dataset indicates that the $0.2\,\%$ bound is also valid in this case.
\end{itemize}

\begin{figure}[!h]
\centering
%\resizebox{\columnwidth}{!}{
\includegraphics[width=.8\columnwidth]{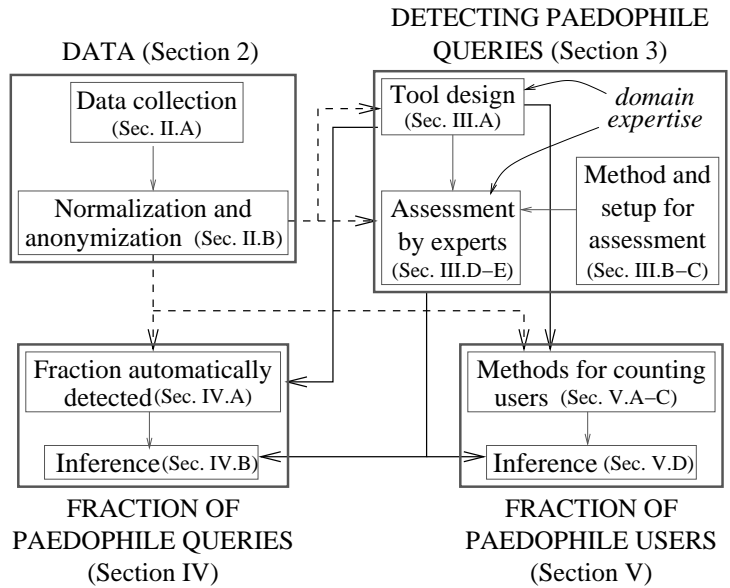}
%}
\caption{
Structure of the paper and main contributions.
}
\label{fig-diagram}
\end{figure}

Figure~\ref{fig-diagram} summarises our global methodology. We describe in Section~\ref{sec-data} our dataset collection and anonymisation. Section~\ref{sec-tool} presents our tool for automatic detection of paedophile queries, our assessment methodology, and the estimates of its error rates by experts of the field. We finally establish the fractions of paedophile queries and users who entered them in Sections~\ref{sec-queries} and~\ref{sec-users}. We discuss related work in Section~\ref{sec-related}.

Finally, we obtain, for the first time at this level of accuracy and reliability, estimates of the two most crucial statistics on paedophile activity in a large P2P system.
We establish reference end-to-end methodology for tackling such issues rigorously. We moreover provide publicly our full datasets, our paedophile query detection tool, and the sets of detected paedophile queries \cite{URL}. We therefore open the way to more research on paedophile activity and other activities in P2P systems, which we discuss in Section~\ref{sec-conclu}.

%%%%%%%%%%%%%%
\section{Data}
\label{sec-data}

Although many extensions exist \cite{wikipediaedonkey}, the \edonkey\ system basically relies on a set of 100 to 300 servers indexing available files and providers for these files. Clients send to these servers keyword-based queries (which may also contain meta-data such as a type of file) describing the content they search for. Servers answer with lists of files matching these keywords (typically, their filenames contain these keywords). Clients may then ask the server for providers of selected files. Once they have obtained this information, they may contact providers directly to obtain the files. Servers only play the role of directories; they do not store any exchanged file, and exchanges take place between clients, from peer to peer. \edonkey\ is currently one of the largest P2P systems used, and this has been true for several years~\cite{schulze09ipoque}.

We collected for this study two independent datasets, in 2007 and 2009. Both consist of a recording of hundreds of millions keyword-based queries received by an \edonkey\ server during a period of time of several weeks. To each query is associated a timestamp and the IP address from which it was received. The 2007 dataset contains in addition the connection port number used for sending each query. Notice that we do not observe exchanges actually occurring between users, and have no access to file content. This is not obtainable in practice at a large scale and is not mandatory for our purpose as we focus on what users seek.

Key features of both datasets are summarised in Table~\ref{tab-data}. We detail the data collection, normalisation and anonymisation procedures below.

\begin{table}[!h] 
\centering 
%\resizebox{\columnwidth}{!}{
\begin{tabular}{|c|c|c|c|c|} 
 \cline{2-5}
\multicolumn{1}{c|}{} & \centering duration & \centering queries & IP addresses & (IP,port) \\
  \hline 
  2007 & 10 weeks & 107,226,021 & 23,892,531 & 50,341,797 \\ 
  \hline 
  2009 & 28 weeks & 205,228,820 & 24,413,195 & {\em n/a} \\ 
  \hline 
\end{tabular}
%}
\caption{Main features of our two datasets after normalisation, anonymisation, and removal of empty queries.} 
\label{tab-data} 
\end{table}

\subsection{Data collection}

We collected the 2007 dataset on one of the main servers at that time, during a 10-week continuous measurement. It consisted in doing an IP-level capture of the UDP traffic on the server. We therefore had to decode this traffic into application-level messages, and then to select keyword-based queries. One advantage of this dataset is that we were able to record both IP addresses and connection ports of clients, which plays a key role in user quantification (Section~\ref{sec-users}). The procedure for this measurement is fully detailed in \cite{tenweeks}.

We collected the 2009 dataset by a 28-week continuous measurement on a medium-sized server at that time.  We activated the log capability embedded in the standard server software, which records directly (in a human-readable textual format) timestamped keyword-based queries and the IP address they were received from. We observe this way both UDP and TCP traffic, but have no access to information on connection ports.

\begin{figure}[!h]
\centering
%\resizebox{\columnwidth}{!}{
%  \begin{fmpage}{0.8\linewidth}
\includegraphics[width=0.8\columnwidth]{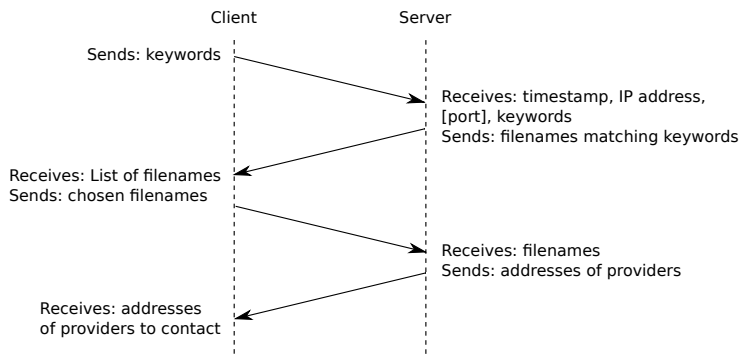}
%  \end{fmpage}
%}
\caption[client-server]{
  Client-server exchanges during an \edonkey\, session.
}
\label{fig-clientserver}
\end{figure}

Figure~\ref{fig-clientserver} presents the roles of the client and server during an \edonkey\, session. Due to the measurement techniques used on the different servers, some information may or may not be available in our collected data.

These datasets come from different servers, with different importance in the network (one of the main servers for 2007 and a medium one for 2009), and the measurement procedures were very different. In addition, three years elapsed between the collection of these two datasets, with many evolutions: users are not the same, the \edonkey\ software and protocols evolved significantly, many servers were stopped due to actions of copyright holders \cite{delahunty06ed2k}, other P2P systems increased in popularity, etc. We therefore consider these two datasets as complementary, which gives some insight on the robustness of our observations to measurement conditions.

\subsection{Normalisation and anonymisation}

The traffic we observe contains much personal information, which we must remove to comply with ethical and legal restrictions. The main such information is the IP address from which each query was made, but text queries also contain personal information: some individuals enter their name or their friends' names, phone numbers, or even credit card numbers as search-strings \cite{adar2007user,tenweeks,allman07issues,narayanan08de}. Finally, users interested in rare contents may enter very specific keywords, usable for identifying them.

Anonymisation of internet traces is a subtle issue in itself \cite{adar2007user,allman07issues,narayanan08de}. The challenge consists in obtaining rich data while fully preserving user privacy. Relying on state-of-the-art knowledge, we set up an appropriate anonymisation procedure which we describe below. We applied it independently to both datasets (as it had to be performed at measurement time).

First notice that anonymising IP addresses with a hash code is not satisfactory: one may decode addresses by applying the function to the $2^{32}$ possible addresses. Likewise, encryption relying on a secret key is unsure on such data. We therefore chose to encode addresses according to their order of appearance in the captured data: we replace the first observed IP address by 0, the second by 1 and so on. We proceed similarly for connection port numbers. This anonymisation is consistent: we always replace a same IP address or port by the same integer. Although computationally expensive, it has three key advantages: it ensures a very strong anonymisation level, it is feasible in real-time during the measurement (which is mandatory), and it makes further use of the dataset much easier.

We apply the following scheme to keyword-based que\-ries. First we replace all accented characters by the corresponding unaccented letter. Then we convert all characters to lower-case, replace all non-alphanumeric characters by a space, and remove successive spaces. We call the obtained queries {\em normalised queries}. They contain only series of alphanumeric characters separated by spaces, which we call {\em words}.

In order to anonymise this normalised data, we have to distinguish between personal (sensitive) and general (non-sensitive) information. We assume that a same word entered by many users (and thus in many queries) is not sensitive \cite{adar2007user,allman07issues}. For instance, a sensitive name or phone number would appear only in a few queries, or in many queries but entered by a same user. We finally remove all words appearing in normalised queries from less than 50 distinct IP addresses, which ensures a very high level of anonymisation.

Numbers, in particular short numbers, tend to appear in many contexts. They therefore appear in clear after the procedure above, which raises anonymisation concerns. In particular, phone numbers often appear as space-separated series of two or three digits, which appear in queries from more than 50 IP addresses and so are not removed. A solution would be to remove all numbers, but keeping such information is crucial here as they often indicate ages in paedophile queries, see Section~\ref{sec-tool-design}. We therefore decided to remove numbers of more than two digits, and numbers with value greater than 16. This hides most numbers while keeping the age information we need.

Finally, short words raise similar concern. In particular, charset encoding problems sometimes lead to strings where a blank space is inserted between any two consecutive characters. Single characters are then considered as words, and they will most probably appear frequently. In the procedure above, this means that the corresponding queries will appear in clear. To avoid this, we remove all words composed of only one letter. Removing words with only two or three letters would also make sense, but this would discard important information, see Section~\ref{sec-tool-design}. Inspection of our data showed that keeping them raises no significant anonymisation concern. 

Figure~\ref{fig-anonqueries} presents two examples of queries, and their counterparts after the normalisation and anonymisation procedures.

\begin{figure}[!h]
%\centering
%  \begin{fmpage}{0.8\linewidth}
%{\em Ann O'Nymous lives 87 Zenoria St London}
{\em Photo Scarlett O'Hara \'et\'e ``Champs-\'Elys\'ees'' }\\
$\hookrightarrow$ {photo scarlett o hara ete champs elysees}\\
$\hookrightarrow$ {photo scarlett hara ete champs elysees}\\

\vskip0.2cm

{\em Credit card 1234 4567 7654 4321 johndoe7643@something.com}\\
$\hookrightarrow$ {credit card 1234 4567 7654 4321 johndoe7643 something com}\\
$\hookrightarrow$ {credit card something com}\\
%\end{fmpage}

\caption{ Two queries presented in their raw, normalised and anonymised
  versions, from top to bottom. Raw queries are italicised. \newline In the
  first query, accented letters are first replaced by their non-accented
  counterparts. All letters are transformed into lower-case, the apostrophe and
  quotation marks are replaced by spaces. We obtain a {\em normalised query}.
  The subsequent anonymisation procedure only removes the remaining {\em o}, now
  a single-letter word. \newline In the second query, the normalisation consists
  of putting the text in lower-case and replacing the arobace and dot by spaces.
  The anonymisation then removes the 4-digit groups, and {\em johndoe7643}, a
word not sufficiently common to appear in queries from more than 50 distinct IP
addresses. Consecutive spaces are eventually trimmed. } 

\label{fig-anonqueries} 
\end{figure}

\section{Detecting paedophile queries}
\label{sec-tool}

In this section we design a tool for automatic identification of paedophile queries in large sets of queries, most of which are not paedophile. The standard way for doing so is to use machine learning approaches, which rely on the prior knowledge of a representative set of paedophile queries. No such dataset is currently available, though, and constructing it without an automatic tool would require expert inspection of huge sets of queries, which is by far too time-consuming and costly.

We therefore rely on domain knowledge of paedophile keywords and ad hoc observations to manually design our tool (Section~\ref{sec-tool-design}). Such a tool is necessarily prone to errors: some paedophile queries may not be tagged as such, and some non-paedophile queries may be tagged as paedophile. It is therefore crucial to obtain precise estimates of our error rates in order to make quantification of paedophile activity possible. This raises specific challenges in our context, which we address in Section~\ref{sec-tool-perf}. We then set up an assessment framework which we submit to several independent and highly qualified experts (Section~\ref{sec-tool-setup}). Using the results of this assessment (Section~\ref{sec-tool-experts}), we finally obtain reliable estimates of our tool's error rates (Section~\ref{sec-tool-results}).

\subsection{Tool design}
\label{sec-tool-design}

Our tool for detecting paedophile queries consists in performing a series of simple lexical tests (matchings of keywords in queries), each aimed at detecting paedophile queries of a specific form.
We built a first set of rules based on our expertise in the paedophile context acquired for several years of work on the topic with law-enforcement personnel \cite{URL}. We then manually inspected the results, identified some errors, and corrected them by adding minor variants to these general rules. We iterated this until obtained improvements became negligible.

\begin{figure}[!h]
\centering
%\resizebox{\columnwidth}{!}{
\includegraphics[width=\columnwidth]{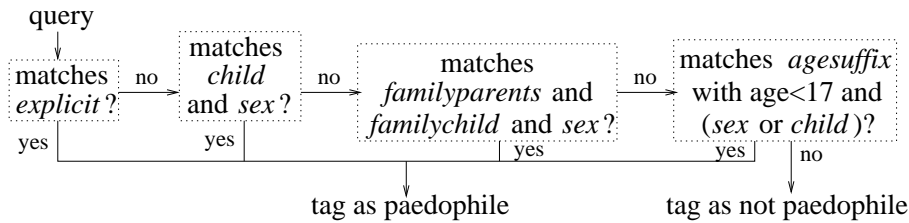}
%}
\caption[short version to allow cite]{
Sequence of tests performed by our tool. Each matching consists in detecting if the query contains words from specific sets (named {\em explicit}, {\em child}, {\em sex}, {\em familyparents}, {\em familychild}, and {\em agesuffix}). See \cite{URL} for these sets of keywords and the tool source code.
}
\label{fig-tool}
\end{figure}

We describe out final rules below, and outline the detection steps in Figure~\ref{fig-tool}.

According to experts of paedophile activity, some keywords point out exclusively such activity in P2P systems, \ie{} they have no other meaning and are dedicated to the search of paedophile content. Typical examples include {\em qqaazz}, {\em r@ygold}, or {\em hussyfan}.  We therefore built a list of specific keywords, called {\em explicit}, and we tag any query containing at least one word from this list as paedophile.

Many paedophile queries contain words related to children or childhood and words related to sexuality, such as {\em child} and {\em sex}.  We therefore constructed a list of keywords related to childhood, called {\em child}, and a list of keywords related to sexuality, called {\em sex}. We tag any query containing a keyword in both lists as paedophile.  Notice that this may be misleading in some cases, for instance for queries like {\em destinys child sexy daddy} (a song descriptor).

A variant of this rule, which we added to the two previous ones, consists in tagging as paedophile the queries containing words related to family, denoting parents {\em and} children (stored in two lists called {\em familyparents} and {\em familychild}), and a word from the {\em sex} list.

Finally, many queries contain age indications under the form {\em n yo}, generally meaning that the user is seeking content involving {\em n years old} children. Other suffixes also appear in place of {\em yo}: {\em yr}, {\em years  old}, etc. We identified such suffixes and built a list named {\em agesuffix}. Age indications are strong indicators of paedophile queries, but they are not sufficient in themselves: they also occur in many non-paedophile queries (for example when the user seeks a computer game for children). We decided to tag a query as paedophile if it contains age indication lower than 17 (greater ages appear in many non-paedophile queries) {\em and} a word in the {\em sex} or {\em child} lists.

In all situations above, although most keywords are in English, local language variations occur, in particular French, German, Spanish, and Italian versions. A few queries in rarer languages, such as Russian and Chinese, also occur. We included the most frequent translations in our sets of keywords.

We provide the exact rules implemented in our tool (including the sets of keywords we use) and the tool itself at \cite{URL}.

\subsection{Method for tool assessment}
\label{sec-tool-perf}

\begin{figure}[!h]
\centering
%\resizebox{\columnwidth}{!}{
\includegraphics[width=\columnwidth]{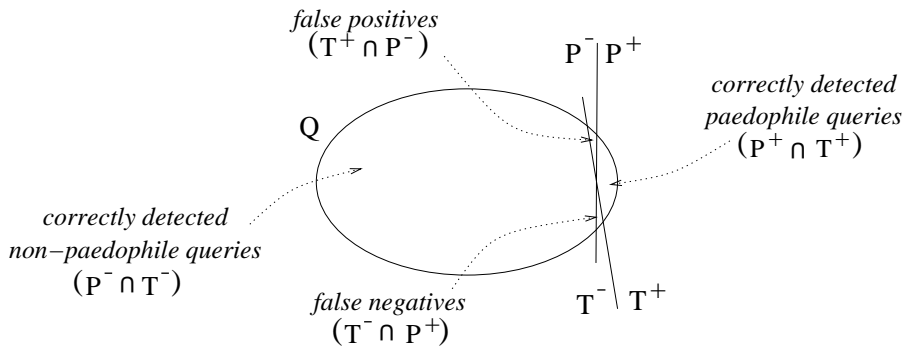}
%}
\caption{
Illustration of our notations. The ellipse represents the set of all queries, $Q$. The vertical line labelled $P\moins/P\plus$ divides $Q$ into the set of non-paedophile queries $P\moins$ (left) and the set of paedophile queries $P\plus$ (right). Likewise, the vertical line labelled $T\moins/T\plus$ divides $Q$ into the set of queries tagged as non-paedophile by the tool, $T\moins$ (left), and the set of queries it tags as paedophile, $T\plus$ (right).
}
\label{fig-notations}
\end{figure}

Let us consider a set $Q$ of queries, and let us denote by $P\plus$ (resp. $P\moins$) the set of paedophile (resp. non-paedophile) queries in $Q$. Let us denote by $\tagged\plus$ (resp. $\tagged\moins$) the subset of $Q$ which is tagged as paedophile (resp. non-paedophile) by our tool. Figure~\ref{fig-notations} provides an illustration of our notations.

Ideally, we would have $\tagged\plus = P\plus$, which would mean that our tool makes no mistake. In practice, though, there are in general paedophile queries which our tool mis-identifies, \ie\ queries in $\tagged\moins \cap P\plus$. Such queries are called {\em false negatives} (the tool produces an erroneous negative answer for them). {\em False positives}, \ie\ queries in $\tagged\plus \cap P\moins$, are defined dually. These notions are classical in data mining, see for instance \cite{foss08estimating}.

The numbers of false positives and false negatives describe the performance of our tool on $Q$. Notice however that they strongly depend on the size of $P\plus$ and $P\moins$. In our situation, we expect $P\plus$ to be much smaller than $P\moins$ (most queries are not paedophile), which automatically leads to small numbers of false negatives, even in the extreme (and useless) case where the tool would give only negative answers.

To evaluate the performance of a tool in such situations, two natural notions of false positive and false negative rates coexist. Both will prove to be useful here.

\medskip

First, one may consider the false negative (resp. positive) rate when all inspected queries are paedophile (resp. non-paedophile):
$$
f\moins = \frac{|\tagged\moins \cap P\plus|}{|P\plus|}
\mbox{\ \ \ \ and\ \ \ \ }
f\plus = \frac{|\tagged\plus \cap P\moins|}{|P\moins|}.
$$

An estimate of $f\plus$ may then be obtained by sampling a random subset $X$ of $P\moins$ (\ie\ random non-paedophile queries) and manually inspecting the results of the tool on $X$. Constructing $X$ is easy: as most queries are non-paedophile, one may sample random queries and then manually discard the ones which are paedophile. As long as $X$ is small, this has a reasonable cost.  However, the fraction of queries in $X$ which will be tagged as paedophile by our tool will be extremely small. As a consequence, an estimate of $f\plus$ obtained this way would be of poor quality.

Conversely, an estimate of $f\moins$ may be obtained by sampling a random subset $X$ of $P\plus$ (\ie\ random paedophile queries) and manually inspecting the results of the tool on $X$. As $P\plus$ is very small and unknown, sampling $X$ is a difficult task. We may however approximate it using the notion of {\em neighbour} queries as follows.

Given a query $q$ in $Q$, its {\em backward neighbour} is the last query in $Q$ which was received from the same IP address as $q$ less than two hours {\em before} $q$, if it exists\,\footnote{We chose this threshold after examining the distributions of query interarrival times; it must be large enough to lead to many cases where neighbour queries exist, while being small enough to make it probable that neighbours of a query are related to this query. To this regard, two hours is a good compromise (which we confirm in Section~\ref{sec-tool-experts}, Table~\ref{tab-classif}), but a wide range of values around this specific value lead to similar observations. The value we chose for the threshold is consistent with the conclusions of \cite{churnMagnienBenamara}, a study dedicated to inter-query times in \edonkey.}. We therefore expect it was entered by the same user as $q$, seeking the same kind of content\,\footnote{IP addresses are not enough to distinguish between users (see Section~\ref{sec-users}) but {\em many} neighbours of paedophile queries are themselves paedophile (see Section~\ref{sec-tool-experts}, Table~\ref{tab-classif}), which is what we need.}. Likewise, we define the {\em forward neighbour} of $q$ as the first query in $Q$ which was received from the same IP address as $q$ within two hours {\em after} $q$.

We denote by $N(q)$ the set containing the backward and forward neighbours of a query $q$. This set may be empty, and contains at most two elements. We denote by $N(S) = \cup_{q\in S} N(q)$ the set of neighbour queries of all queries in set $S$, for any $S$. We guess that queries in $N(P\plus)$, \ie\ the neighbours of paedophile queries, are also paedophile with high probability (much higher than random queries in $Q$). We expect this to be also true for queries in $N(\tagged\plus)$, which is confirmed in Section~\ref{sec-tool-experts}, Table~\ref{tab-classif}.

Obviously, $N(\tagged\plus) \cap P\plus \subseteq P\plus$, but $N(\tagged\plus) \cap P\plus \not\subseteq \tagged\plus$ in general. In other words, $N(\tagged\plus)$ probably contains queries in $P\plus$ (\ie\ paedophile queries) which are {\em not} detected by our tool. If we consider the queries in $N(\tagged\plus) \cap P\plus$ as random paedophile queries, then they may be sampled to construct a set $X$ of random paedophile queries suitable for estimating $f\moins$. As $X$ contains only paedophile queries, this estimate is equal to the number of  queries in $X$ not detected as paedophile by our tool divided by the size of $X$.

Notice that the queries in $X$ may actually be biased by the fact that they are derived from $\tagged\plus$: the probability that a user enters a paedophile query which the tool is able to detect is higher if this user already entered one such query (he/she may enter in both cases keywords detected by our tool). As a consequence, our estimate of $f\moins$ may be an under-estimate.

\medskip

Finally, one cannot, in our context, evaluate $f\plus$ properly; on the
contrary, we are able to give a reasonable \mbox{(under-)}estimate for $f\moins$. But both $f\plus$ and $f\moins$ are needed to evaluate the performance of our tool.

\medskip

In order to bypass this issue, we consider the following variants of false negative and false positive rates, which capture the probability that the tool gives an erroneous answer when it gives a positive (resp. negative) one:
$$
f\primeplus = \frac{|\tagged\plus \cap P\moins|}{|\tagged\plus|}
\mbox{\ \ \ \ and\ \ \ \ }
f\primemoins = \frac{|\tagged\moins \cap P\plus|}{|\tagged\moins|}.
$$

An estimate of $f\primeplus$ may be obtained by sampling a random subset $X$ of $\tagged\plus$ (\ie\ a random set of queries for which our tool gives a positive answer) and by manually inspecting this subset in order to obtain the number of false positives.  We expect all sets involved in these computations to be of significant size (which is confirmed in Section~\ref{sec-tool-experts}), so there is no obstacle in computing a reasonable estimate for $f\primeplus$.

Conversely, an estimate of $f\primemoins$ may be obtained by sampling a random subset $X$ of $\tagged\moins$ and inspect it to determine the number of false negatives, \ie\ the number of queries in $X$ which actually are paedophile. However, as paedophile queries are expected to be very rare, the number of observed false negatives will be extremely small as long as $X$ is of reasonable size.

\medskip

Therefore, one may easily obtain a significant estimate of $f\primeplus$, but computing a reasonable estimate for $f\primemoins$ is not tractable in our case.

\medskip

Finally, the quantities we will use for evaluating the quality of our tool are $f\primeplus$ (the rate of errors when our tool decides that a query is paedophile) and $f\moins$ (the rate of paedophile queries that our tool mis-classifies as non-paedophile), which we are able to properly estimate. We describe our practical procedure for computing these estimates in the following sections and provide the obtained estimates in Section~\ref{sec-tool-results}.

%%%%%%%%%%%%%%%%%%%%%%%%%%%%%
\subsection{Assessment setup}
\label{sec-tool-setup}

In order to apply the method for quantifying our tool quality described above, we need to identify actual paedophile queries in some specific sets. To do so, we resort to independent experts of paedophile activity who manually inspect and tag these queries. We describe here the construction of these sets, the experts who helped us, and the interface we provided to them.

\subsubsection*{Query selection}

Because the 2009 dataset was not yet available when we designed our tool and assessed it, we used the 2007 dataset for sampling queries to assess. We denote by $Q$ the whole set of queries, and use the formalism of Section~\ref{sec-tool-perf}. We divide $Q$ into three sets (with overlap): $\tagged\moins$ (queries tagged as not paedophile by our tool), $\tagged\plus$ (queries it tagged as paedophile), and $N(\tagged\plus)$ (neighbours of queries it tagged as paedophile). These three sets are easy to compute from $Q$ using our tool.

Notice that some queries in $\tagged\plus$, \ie\ some queries which are tagged as paedophile by the tool, are composed of only one word.
Then, this word is necessarily a word in the {\em explicit} paedophile keywords list described in Section~\ref{sec-tool-design}. These keywords are known to have a very strong paedophile nature. Therefore, if such a keyword appears alone in a query, then this query surely is paedophile. We therefore increase the efficiency of our assessment by not submitting these one-keyword queries to experts. We denote by $\tagged\plus_1$ the set of queries in this set, and by $\tagged\plus_{>1}$ the queries in $\tagged\plus$ composed of more than one word. Our optimisation consists in using the fact that $\tagged\plus_1 \subseteq P\plus$, and so use only $\tagged\plus_{>1}$ for assessment.

We finally construct the sets of queries to assess by selecting $1,000$ random queries in each of the sets $\tagged\moins$, $\tagged\plus_{>1}$ and $N(\tagged\plus)$ (thus $3,000$ queries in total\,\footnote{Since there is an overlap between $N(\tagged\plus)$ and the other sets, we could have sampled some queries more than once, leading to less than $3,000$ queries in total.  Since $1,000$ is small compared to the total sizes of the three sets, this did not happen here.}). This leads to three subsets which we denote by $\overline{\tagged\moins}$, $\overline{\tagged\plus_{>1}}$, and $\overline{N(\tagged\plus)}$ respectively. Notice that carefully tagging $3,000$ queries already is a heavy task for experts. For this reason, we did not reproduce the assessment on the 2009 dataset and simply checked manually that its outcome would be very similar.

\subsubsection*{Experts}

Once we selected sets of queries for which we need expert classification, the choice of experts is a crucial step. Indeed, deep knowledge of online paedophile activity is needed, if possible with a focus on P2P activity and/or query analysis. Such expertise is extremely rare, even at the international level. It is present mainly in law-enforcement institutions, where special units are devoted to fighting (online) paedophile activity, and in NGOs dedicated to similar tasks (but with a different approach, in general). Some security consultants also have this kind of knowledge.

Thanks to our involvement in international research projects on paedophile activity for several years, with partners in various law-enforcement agencies and NGOs in several countries, we were able to contact a large number of specialists who may play the role of experts in our study. We were for instance able to send a call for experts on the main international mailing-list of law-enforcement personnel working on cybercrime.

We finally obtained a set of 21 volunteers for participating to our assessment task. These participants are personnel of various law-enforcement institutions (including Europol and the main French and Danish national agencies) and well-established NGOs (including the {\em National Center for Missing \& Exploited Children}, {\em Nobody's Children Foundation}, {\em Action Innocence Monaco} and the {\em International Association of Internet Hotlines}). A few security consultants also contributed.

We later conducted an assessment of participants themselves to ensure that we use only answers from relevant experts, see Section~\ref{sec-tool-experts}.

\subsubsection*{Interface}

We set up a web interface to make it convenient for participants to tag queries.  All $3,000$ queries were presented in a different random order for each participant, thus avoiding possible bias due to a specific order. Moreover, it was possible for participants to tag only a part of the $3,000$ proposed queries, thus allowing them to contribute even if they had limited time.

We proposed five possible answers for each query: {\em paedophile}, {\em probably paedophile}, {\em probably not paedophile}, {\em not paedophile}, and {\em I don't know}. To help participant's choice, we displayed each query with its backward and forward neighbours (defined in Section~\ref{sec-tool-perf}), when they existed. This was of great help in tagging ambiguous queries.

%%%%%%%%%%%%%%%%%
\subsection{Expert results}
\label{sec-tool-experts}

The answers collected from our 21 participants are summarised in Table~\ref{tab-expert-results}. Each of them tagged more than 300 queries (\ie\ $10\,\%$ of the whole), and 12 tagged more than 2,000.

\begin{table}[!h]
\centering
\resizebox{\columnwidth}{!}{\begin{tabular}{|c|c|c|c|c||c||c|}
\cline{2-5}
\multicolumn{1}{c|}{} & \emph{prob.} & \emph{don't} & \emph{prob.} & \emph{not} &\multicolumn{2}{c}{}\\
\cline{1-1}
\cline{6-7}
\emph{paedo} & \emph{paedo} & \emph{know} & \emph{not} & \emph{paedo} & total & relevance\\
\hline
1530&149&25&66&1230&3000&99.6 \\
1381&247&125&580&667&3000&98.5 \\
1679&89&2&113&1117&3000&99.2 \\
1603&201&99&174&923&3000&99.3 \\
1598&5&15&1&1381&3000&98.9 \\
128&81&1&26&124&360&96.3 \\
216&154&0&142&132&644&98.6 \\
1624&126&16&165&581&2512&99.7 \\
351&16&2&16&27&412&99.6 \\
647&119&71&40&439&1316&98.9 \\
1174&111&20&64&789&2158&99.3 \\
335&17&1&70&166&589&97.1 \\
641&383&4&112&753&1893&96.6 \\
1071&546&2&453&928&3000&87.3 \\
1554&197&28&327&894&3000&98.2 \\
305&270&24&89&181&869&98.3 \\
371&1017&496&570&546&3000&95.7 \\
976&936&405&594&89&3000&95.5 \\
344&12&10&70&156&592&99.0 \\
845&139&323&175&182&1664&98.1 \\
1506&120&6&25&393&2050&98.3 \\
\hline
\end{tabular}}
\caption{Assessement results for each participant. Each line corresponds to a participant, and gives the number of answers of each kind he/she provided, his/her total number of answer, and his/her relevance.}
\label{tab-expert-results}
\end{table}

\subsubsection*{Expert selection}

Despite our efforts to select appropriate contributors, some may have an inadequate knowledge of our particular context (paedophile queries in a P2P system), and lower the quality of our results by entering erroneous answers. In order to identify such cases, we examined the answers of each participant to the queries which contain an {\em explicit} paedophile keyword, \ie\ a word in our {\em explicit} list (defined in Section~\ref{sec-tool-design}). As already said, these keywords are well acknowledged paedophile keywords, which all experts of the field consider as strong indicators of paedophile queries.

The set of all queries submitted to contributors contains 1,003 queries which contain at least one explicit paedophile keyword. We provide in Table~\ref{tab-expert-results} (rightmost column) the percentage of these queries which the corresponding contributor tagged as {\em paedophile} or {\em probably paedophile}. For all contributors except one, this percentage is above $95\,\%$, thus showing that these contributors recognise these keywords. The remaining contributor only slightly disagrees with a ratio of $87.3\,\%$.

The ratios discussed above may be misleading if a contributor tags all or almost all queries as paedophile. Table~\ref{tab-expert-results} gives precise insight on this. The answers of most contributors are well balanced between all possible answers, except for three contributors (see for instance the last line of Table~\ref{tab-expert-results}). Manual inspection shows that these contributors focused preferentially on paedophile queries (they did not tag all queries), which does not invalidate their answers. We therefore keep them in our expert set.

Finally, we obtain 42,059 answers provided by 21 experts who contributed at least 300 answers each. This leads to an average of slightly more than 14 experts assessing each query, which is sufficient for our purpose.

\begin{table}[!h]
\centering
%\resizebox{\columnwidth}{!}{
\begin{tabular}{|l|c|c|c|}
\cline{2-4}
\multicolumn{1}{c|}{} & \multicolumn{3}{|c|}{random subset}\\
\multicolumn{1}{c|}{} & $\overline{\tagged\moins}$ & $\overline{\tagged\plus_{>1}}$ & $\overline{N(\tagged\plus)}$\\
\hline
{\em paedophile} & 63 & 11,530 & 8,286 \\
{\em probably paedophile} & 237 & 2,303 & 2,395 \\
{\em I don't know} & 1,009 & 208 & 458 \\
{\em probably not paedophile} & 2,294 & 336 & 1,242 \\
{\em not paedophile} & 9,537 & 241 & 1,920 \\
\hline
Total & 13,140 & 14,618 & 14,301 \\
\hline
\end{tabular}
%}
\caption{
Number of votes of each kind for each considered set.
}
\label{tab-votes}
\end{table}

The distribution of these answers among the queries of each considered set is given in Table~\ref{tab-votes}. It is in accordance with what one would expect if our tool performs well, and if our assumption that $\overline{N(\tagged\plus)}$ should contain many paedophile queries is verified. We analyse this in more details now.

\subsubsection*{Classification of queries}

For each query $q$ submitted to experts in our assessment procedure, we denote by $q\plusplus$ the fraction of experts (among the ones who provided an answer for $q$) which tagged it as {\em paedophile} and by $q\plus$ the fraction of experts which tagged it as {\em paedophile} or {\em probably paedophile}. We define $q\moins$ and $q\moinsmoins$ dually. Notice that $q\plus + q\moins < 1$ in general, as some {\em I don't know} answers were provided (the fraction of such answers is $1-q\plus-q\moins$). Moreover, $q\plus \ge q\plusplus$ and $q\moins \ge q\moinsmoins$ for all $q$.

\begin{figure}[!h]
\centering
\resizebox{\columnwidth}{!}{
\includegraphics[width=\columnwidth]{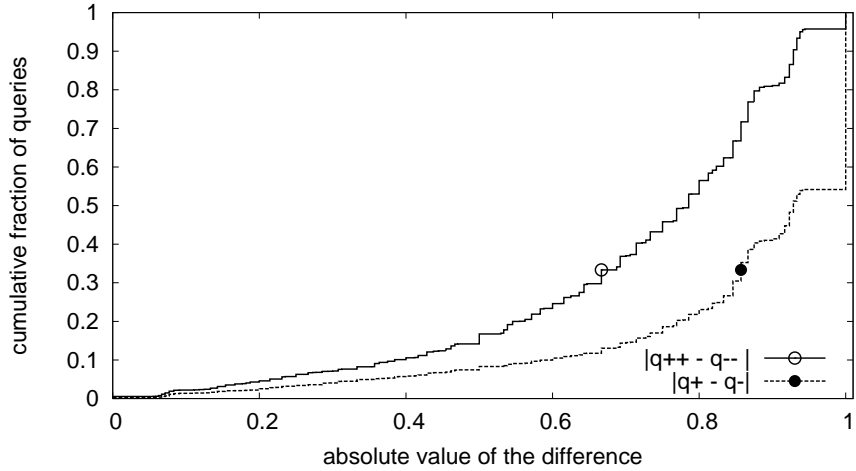}
}
\caption{
The cumulative distribution (CDF) of the absolute value of differences between $q\plusplus$ and $q\moinsmoins$ and between $q\plus$ and $q\moins$ for each query: a point at coordinate $(x,y)$ means that a fraction $y$ of queries have a difference lower than $x$.
}
\label{fig-diff}
\end{figure}

In order to classify queries according to expert answers, we expect to observe that each query $q$ has either a high $q\plus$ (resp. $q\plusplus$) or a high $q\moins$ (resp. $q\moinsmoins$), but not both or neither, meaning that experts agree on the nature of $q$.  Figure~\ref{fig-diff} displays the difference between $q\plus$ and $q\moins$ and between $q\plusplus$ and $q\moinsmoins$ for all queries. These plots grow very slowly for small values on the horizontal axis, showing that only very few queries have a small difference. On the contrary, for many queries, the difference is very large: above $0.8$ for $1,305$ queries (over $3,000$) in the case of $q\plusplus$ and $q\moinsmoins$, and for $2,308$ queries in the case of $q\plus$ and $q\moins$.

% Figure~\ref{fig-corr} gives much insight on this. It displays the correlations between these values. Queries on which many experts disagree are close to the diagonal (the fraction of experts considering them as paedophile is close to the fraction of experts considering them as non-paedophile). We observe that only few queries are in this situation. Notice that dots far from the diagonal (in particular in the upper left corner and in the lower right one) actually represent many queries with the same ratios; in such situations the dots overlap.

Only $41$ queries have a difference $|q\plus - q\moins|$ smaller than or equal to $0.1$, which already is significant. Moreover, this number increases very slowly when the difference grows. We therefore classify a query as paedophile if $q\plus - q\moins > 0.1$ and as non-paedophile otherwise. We finally obtain the query classification by experts presented in Table~\ref{tab-classif}.

\begin{table}[!h]
\centering
%\resizebox{\columnwidth}{!}{
\begin{tabular}{|l|c|c|c|}
\cline{2-4}
\multicolumn{1}{c|}{} & \multicolumn{3}{|c|}{random subset}\\
\multicolumn{1}{c|}{} & $\overline{\tagged\moins}$ & $\overline{\tagged\plus_{>1}}$ & $\overline{N(\tagged\plus)}$\\
\hline
paedophile queries & 1 & 985 & 754 \\
\hline
non-paedophile queries & 999 & 15 & 246 \\
\hline
\end{tabular}
%}
\caption{
Number of queries classified as paedophile or not by experts for each considered set.
}
\label{tab-classif}
\end{table}

\subsection{Tool assessment results}
\label{sec-tool-results}

Thanks to the assessment results in Table~\ref{tab-classif} and the expressions given in Section~\ref{sec-tool-perf}, we may now compute estimates of the false positive and false negative rates which describe the quality of our tool.

First notice that, as expected, the number of paedophile queries in the set of
queries tagged as non-paedophile by the tool is very low:
$|\overline{\tagged\moins} \cap P\plus| = 1$. As a consequence, approximating
$f\primemoins = \frac{|\tagged\moins \cap P\plus|}{|\tagged\moins|}$ by
$\frac{|\overline{\tagged\moins} \cap P\plus|}{|\overline{\tagged\moins}|} =
\frac{1}{1,000}$ would yield very poor quality result.

The estimate obtained for $f\primeplus$ is of much better quality. It relies on the following expression:
$$
\begin{array}{lcl}
f\primeplus & = & \frac{|\tagged\plus \cap P\moins|}{|\tagged\plus|}\\
\medskip
 & = & \frac{|\tagged\plus_1 \cap P\moins| + |\tagged\plus_{>1} \cap P\moins|}{|\tagged\plus|}\\
\medskip
 & = & \frac{|\tagged\plus_{>1} \cap P\moins|}{|\tagged\plus|}
\end{array}
$$
%$$
%\begin{array}{l}
%f\primeplus \ = \ \frac{|\tagged\plus \cap P\moins|}{|\tagged\plus|}
%            \ = \ \frac{|\tagged\plus_1 \cap P\moins| + |\tagged\plus_{>1} \cap P\moins|}{|\tagged\plus|}
%            \ = \ \frac{|\tagged\plus_{>1} \cap P\moins|}{|\tagged\plus|}
%\end{array}
%$$
(since $\tagged\plus_1 \cap P\moins = \emptyset$, because all queries in $\tagged\plus_1$ are paedophile, see Section~\ref{sec-tool-setup}).

An estimate of $|\tagged\plus_{>1} \cap P\moins|$ is given by $|\overline{\tagged\plus_{>1}} \cap P\moins| \cdot \frac{|\tagged\plus_{>1}|}{|\overline{\tagged\plus_{>1}}|}$ which leads to:
$$
\begin{array}{lcl}
f\primeplus & \sim & \frac{|\overline{\tagged\plus_{>1}} \cap P\moins|}{|\tagged\plus|} \cdot \frac{|\tagged\plus_{>1}|}{|\overline{\tagged\plus_{>1}}|}\\
\medskip
 & = & \frac{15}{207,340} \cdot \frac{192,545}{1,000}\\
\medskip
 & \sim & 1.39\,\%.
\end{array}
$$

The quality of this estimate is good not only because $|\overline{\tagged\plus_{>1}} \cap P\moins| = 15$ is significant, but also because we evaluate it using a sample of queries in $\tagged\plus_{>1}$, which is much (more than 500 times) smaller than $\tagged\moins$, involved in the estimate of $f\primemoins$.

\medskip

Conversely, the assessment results confirm that estimating $f\plus = \frac{|\tagged\plus \cap P\moins|}{|P\moins|}$ with our data would yield poor quality approximate, as $|\tagged\plus \cap P\moins|$ is small (there are very few paedophile queries), as well as the sample size, compared to the size of $P\moins$.

It is possible to estimate $f\moins$ much more accurately:
$$
\begin{array}{lcl}
f\moins & = & \frac{|\tagged\moins \cap P\plus|}{|P\plus|}\\
\medskip
 & \gtrsim & \frac{|\tagged\moins \cap (\overline{N(\tagged\plus)}\cap P\plus)|}{|\overline{N(\tagged\plus)} \cap P\plus|}\\
\medskip
 & = & \frac{185}{754}\\
\medskip
 & \sim & 24.5\,\%.
\end{array}
$$

This value however is an under-estimate, because we assessed neighbours of detected paedophile queries instead of random paedophile queries. It is equal to the probability that our tool erroneously tags such a neighbour as non-paedophile. There is no {\em a priori} reason to suppose that this leads to huge differences, though, and we therefore expect this bound to be reasonably tight. We will handle this with care in the following.

%%%%%%%%%%%%%%%%%%%%%%%%%%%%%%%%%%%%%%%%
\section{Fraction of paedophile queries}
\label{sec-queries}

In this section, we estimate the fraction of paedophile queries in our two datasets, \ie\ $\frac{|P\plus|}{|Q|}$ for each $Q$ (we use the notations defined in Section~\ref{sec-tool-perf}, the dataset under concern being given by the context). This may be done by sampling a random subset of $Q$ and then submit the queries it contains to experts able to decide whether they are paedophile or not. As we expect that $P\plus$ is very small compared to $Q$ (the fraction of paedophile queries is low), though, this is not feasible in practice: the size of a random set large enough to contain a representative number of paedophile queries is prohibitive for manual inspection.

We therefore use here the automatic paedophile query detection tool designed in Section~\ref{sec-tool}, for which precise information on its error rates is available. We first estimate the fraction of queries in $Q$ tagged as paedophile by the tool, and then infer from it an estimate of the fraction $\frac{|P\plus|}{|Q|}$.

\subsection{Fraction of automatically detected queries}
\label{sec-queries-tagged}

The automatic paedophile query detection tool divides $Q$ into two disjoint subsets: $\tagged\plus$, the set of queries tagged as paedophile by the tool; and $\tagged\moins$, the set of queries tagged as non-paedophile. We estimate here the fraction of queries tagged as paedophile, \ie\ $\frac{|\tagged\plus|}{|Q|}$, in both datasets.

This may be trivially obtained by computing the set $\tagged\plus$ of queries tagged as paedophile by the tool, and then divide it by the total number of queries. We obtain this way ratios slightly above $0.19\,\%$ for both datasets. In order to ensure the relevance of this estimate, though, we go into details below.

\begin{figure}[!h]
\centering
\resizebox{\columnwidth}{!}{
\includegraphics[width=\columnwidth]{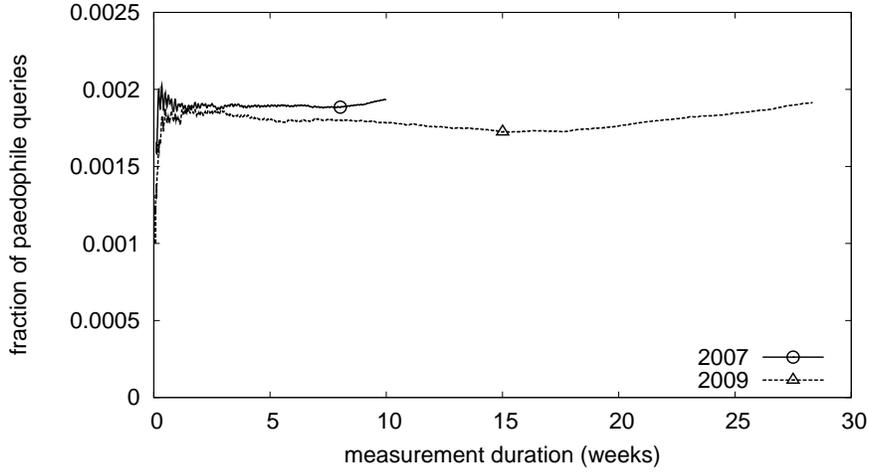}
}
\caption{
Fraction of paedophile queries detected in our datasets as a function of the measurement duration. 
}
\label{fig-queries}
\end{figure}

We first check that the measurement duration is large enough by plotting the fraction of queries tagged as paedophile as a function of the measurement duration, see Figure~\ref{fig-queries}. It clearly shows that this fraction converges rapidly to a reasonably steady value, slightly below $0.2\,\%$; changing this value significantly would need a drastic change in the data.

\begin{figure}[!h]
\centering
\includegraphics[width=.5\columnwidth]{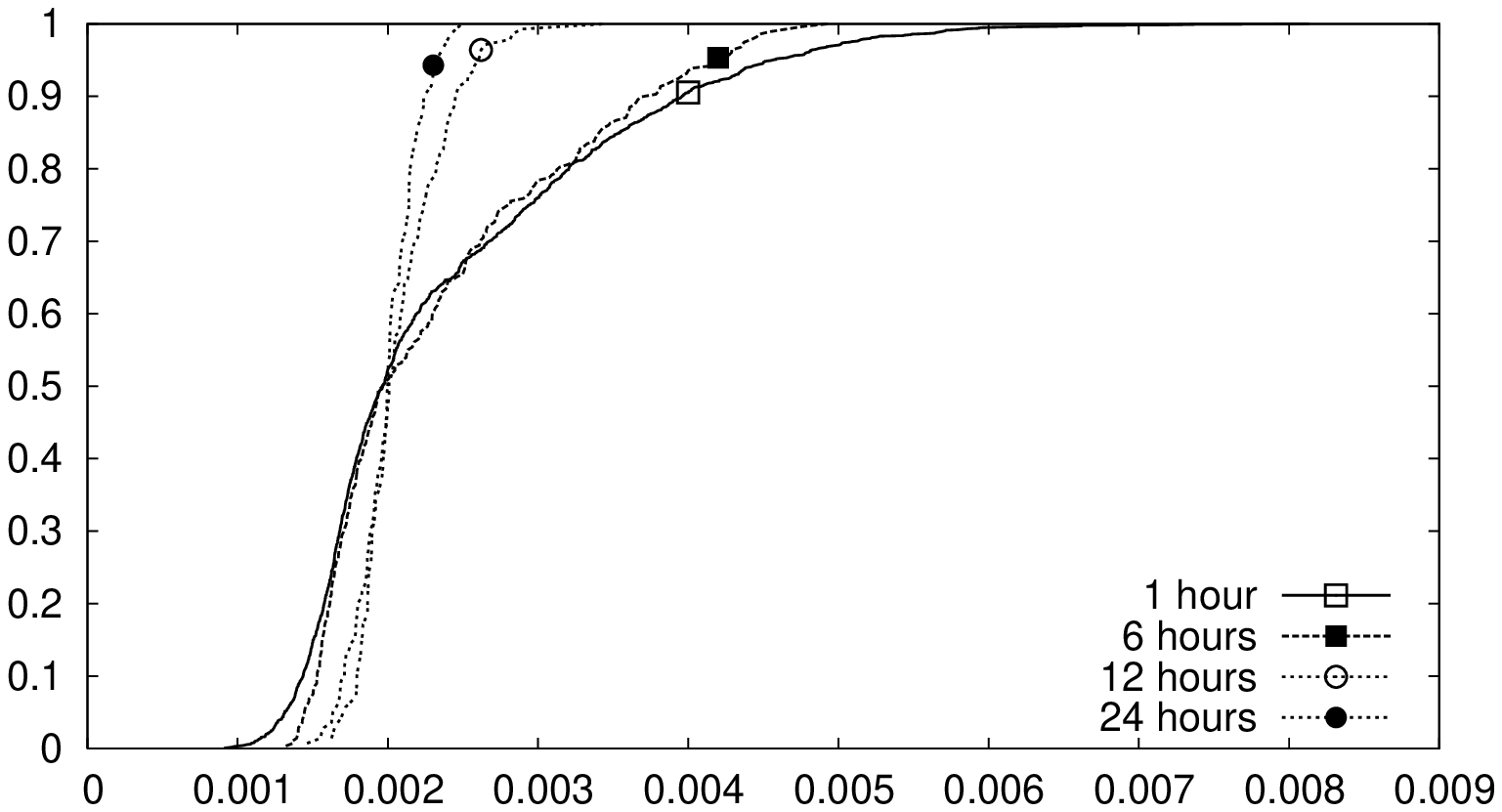}%
\includegraphics[width=.5\columnwidth]{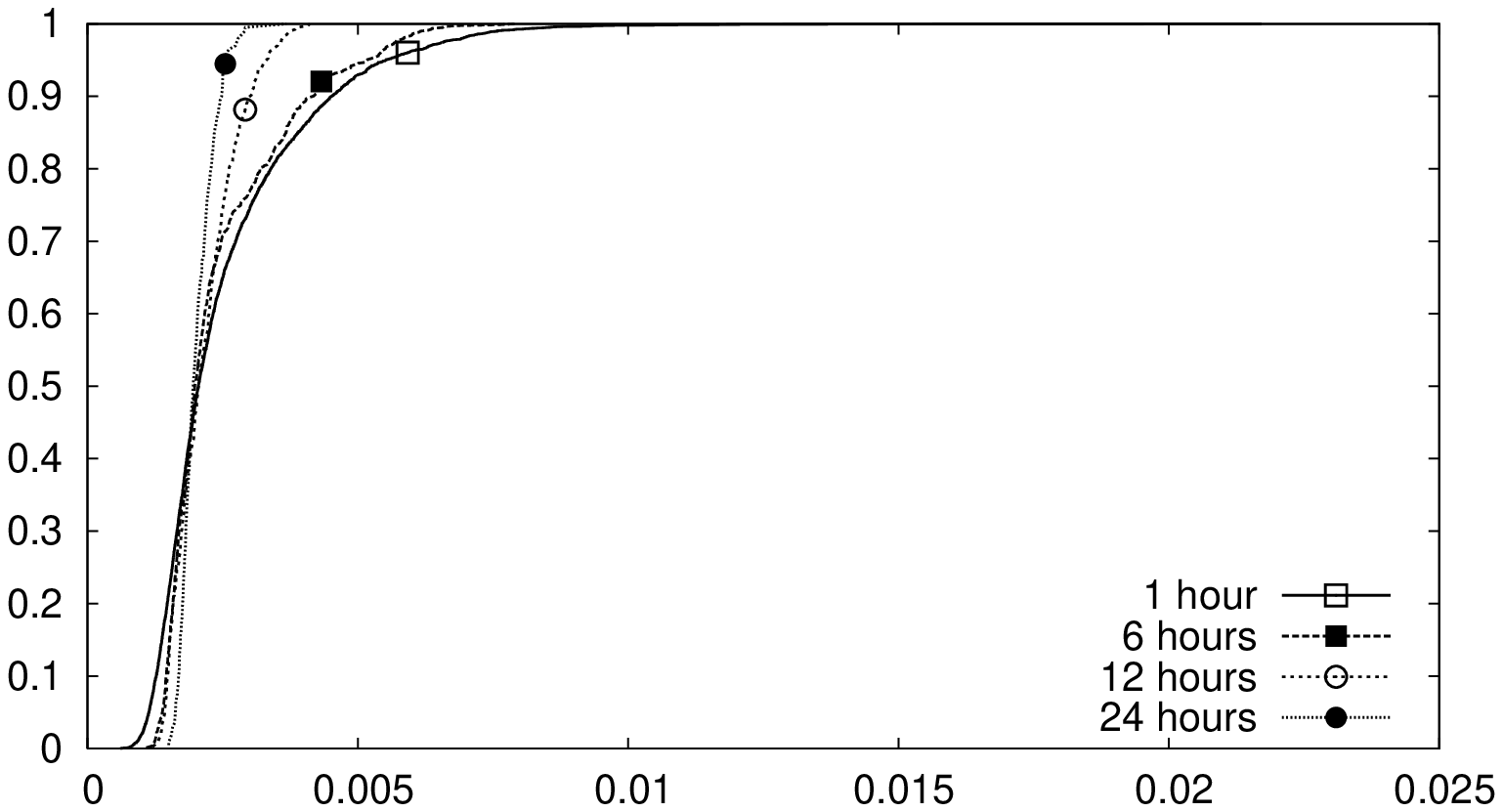}
\caption{
Cumulative distribution of the fraction of paedophile queries observed in time-slices of 1, 6, 12 and 24 hours (each plot corresponds to a size of time slice). A point at coordinates $(x,y)$ means that we observed a fraction $y$ of slices with less than a fraction $x$ of paedophile queries. A sharp vertical increase around $x$ therefore indicates that many slices were observed with a fraction of paedophile queries close to $x$.  }
\label{fig-queries-hour}
\end{figure}

Going further, we plot in Figure~\ref{fig-queries-hour} the cumulative distribution of the fraction of queries tagged as paedophile in all relevant one-hour, 6-hour, 12-hour and 24-hour slices of the measurements\,\footnote{Long-term measurements are subject to interruptions (server or network failures of upgrades, for instance). As a consequence, some time slices in the measurements are not significant (no or few queries are captured during these slices), and thus we discard them when we compute slice statistics.}. This clearly shows that there is a notion of {\em normal}, or {\em median} behaviour for each slice size, and that it is quite independent of slice sizes. The averages of these distributions are all close to $0.2\,\%$, in accordance with our previous computations.

Finally, we conclude that the fraction of queries tagged as paedophile by our tool may be approximated by $\frac{|\tagged\plus|}{|Q|} \sim 0.2\,\%$ in both datasets.

\subsection{Inference}
\label{sec-queries-inference}
\label{sec-queries-result}

We established in Section~\ref{sec-tool-results} reliable estimates for $f\moins$ and $f\primeplus$. As a consequence, we have to infer the size of $P\plus$ from these rates, which may be done as follows:
$$
\begin{array}{lcl}
|P\plus| & = & |P\plus \cap \tagged\plus| + |P\plus \cap \tagged\moins|\\
         & = & |\tagged\plus| (1 - f\primeplus) + |P\plus| f\moins
\end{array}
$$
and so
$$
|P\plus| = \frac{|\tagged\plus| (1-f\primeplus)}{1-f\moins}.
$$

Using $f\moins \gtrsim 24.5\,\%$ and $f\primeplus \sim 1.39\,\%$ (Section~\ref{sec-tool-results}), we obtain:
$$
\frac{|P\plus|}{|Q|} \gtrsim 0.25\,\%
$$
for both datasets.

In other words, at least one query over 400 is paedophile in our two datasets.

Notice that taking $f\moins \sim 50\,\%$, which most certainly is a huge over-estimate, leads to a ratio of approximately $0.38\,\%$ paedophile queries. We therefore conclude that the true ratio is not much larger than $0.25\,\%$.

%%%%%%%%%%%%%%%%%%%%%%%%%%%%%%%%%%%%%%%%%
\section{Fraction of paedophile users}
\label{sec-users}

Although the fraction of paedophile queries is of high interest in itself, the key question when quantifying paedophile activity actually is the fraction of paedophile {\em users}, which we define as users who entered at least one paedophile query.

However, identifying a user in an internet-like environment is a challenge in itself \cite{Bhagwan2003,Stutzbach2006}. Any computer is identified by an IP address at a given time, but even this may change and we are unable in general to detect that a same computer has two different addresses at different times and/or that two computers use the same address. In addition, a same user may use several computers, and several users may use the same computer, making identification of users even more challenging.

More precisely, the following situations occur:
\begin{itemize}
\item several computers in a local network are connected to the internet through a gateway or firewall which performs {\em network address translation} (NAT): they all appear to have the IP address of the gateway or firewall, which is responsible for redistributing the traffic coming from the internet (using ports);
\item internet service providers (ISP) may allocate IP addresses dynamically,
  \ie\ allocate different addresses to a same computer when it connects to the internet at different times, and also allocate the same address to different computers during time;
\item in various places where public internet access is provided (internet cafes, parks, libraries, etc.) or at home, different users  (temporarily) have the same address;
\item and dually, a same user may use several computers (at home, at work, in public places, etc.).
\end{itemize}

This makes user identification at a large scale extremely challenging, and even impossible in practice. Notice however that, in our context, what we need is slightly weaker: we need to make the difference between two users in our dataset in order to avoid mixing their queries.

Indeed, mixing the queries of several users would lead to interpret the corresponding series of queries as a unique series, and thus a unique user.  As we consider a user as paedophile as soon as he/she entered one paedophile query, if one of the corresponding users entered paedophile queries, then the whole series is considered as coming from a paedophile user.  Notice that since the overall fraction of users entering paedophile queries is very small, it is very unlikely that two paedophile users are mixed in this way.  Therefore, mixing the queries of several users leads to a decrease of the total number of observed {\em users}, but in general the number of observed {\em paedophile users} stays the same.  This leads to an over-estimate of the fraction of paedophile users. We call this phenomenon {\em pollution}, and we observe this in practice below.

We explore below different approaches to count users who sent paedophile queries in our datasets. First, we show that identifying users with their IP address only is not sufficient, but that considering the pair composed of their IP address and their connection port provides relevant information. We then study the influence of the measurement duration using sliding windows of different lengths: the bias due to dynamic addressing may be controlled with such an approach. Finally, we consider series of queries received from a same IP address with small inter-query times, which we call {\em sessions}. Indeed, the fraction of sessions containing paedophile queries may be considered as an estimate of the fraction of users entering such queries.

\subsection{IP addresses and connection port numbers}

Two pieces of  information in our datasets may lead to distinguish between users: the IP address from which they sent queries, and the connection port number they used. The latter is important: it makes it possible to distinguish between several users in a same local network with a NAT.

Therefore, we consider here two approximations of the notion of user: we first assume that the IP address is sufficient to distinguish between different users, and then that the pair (IP address, connection port) is sufficient. Notice that this last assumption is necessarily better than the previous one, but comparing the two is enlightening.

\begin{figure}[!h]
\centering
\resizebox{\columnwidth}{!}{
\includegraphics[width=\columnwidth]{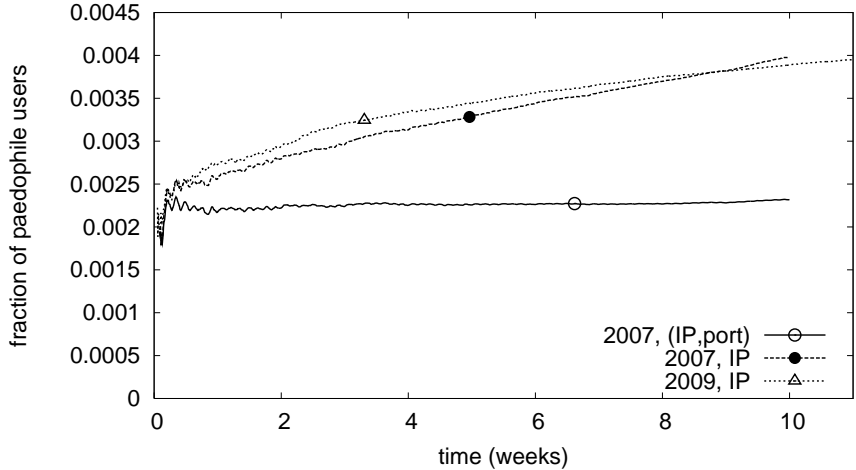}
}
\caption{Fraction of paedophile users detected in our datasets as a function of the measurement duration.}
\label{fig-frac-users}
\end{figure}

We display in Figure~\ref{fig-frac-users} the fraction of IP addresses and (IP,~port) pairs from which at least one paedophile query (as detected by our tool) was entered. We call them paedophile IP and paedophile (IP,~port) pairs for simplification. Notice that only IP addresses are available in the 2009 dataset.

For both datasets, the fraction of paedophile IP addresses clearly grows with the measurement duration.  This reveals the {\em pollution} phenomenon sketched above: as IP addresses may correspond to different users over time, and as one paedophile user is sufficient to make us consider the corresponding address as paedophile, then the probability for any given address to be considered as paedophile grows with measurement time (all IP addresses may eventually be considered as paedophile). This confirms that using IP address alone is misleading in this case.

On the other hand, the fraction of paedophile (IP,~port) pairs in the 2007 dataset has a very different behaviour: it rapidly reaches a steady regime, very similar to the fraction of paedophile queries studied in Section~\ref{sec-queries}, Figure~\ref{fig-queries}. This shows that pollution due to dynamic allocation of addresses and ports is negligible in this case.

We finally conclude that the fraction of paedophile (IP,~ports) pairs is meaningful, and that this fraction is slightly above $0.22\,\%$ here.

\subsection{Varying measurement duration}
\label{subsec-window}

Figure~\ref{fig-frac-users} shows that increasing the measurement duration leads to an increase of the pollution of IP addresses by paedophile users.  Therefore, considering shorter measurement windows leads to a better handling of the pollution phenomena.  On the other hand, this leads to less observed data, and  therefore to less reliable results.

\begin{figure}[!h]
\centering
\resizebox{\columnwidth}{!}{
\includegraphics[width=\columnwidth]{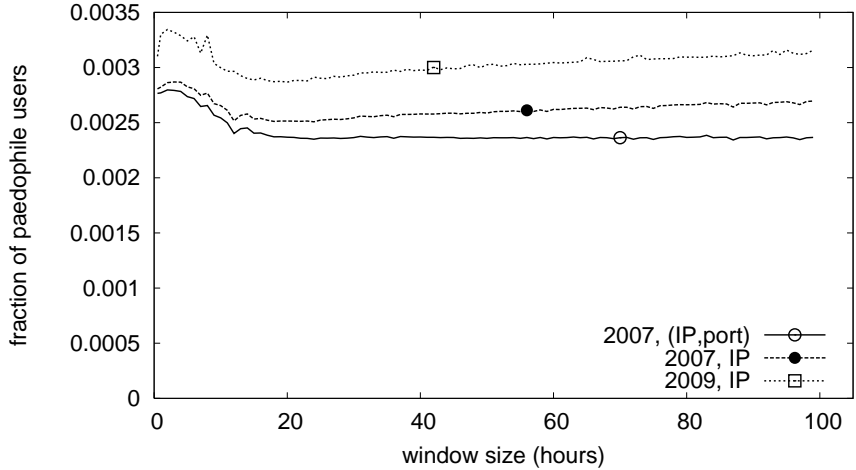}
}
\caption{
Fraction of paedophile users as a function of the measurement window size.
}
\label{fig-window-users}
\end{figure}

To study this, we divide our datasets into small measurement windows, and compute the observed fraction of paedophile IP addresses or (IP,~port) pairs for all windows. The distribution of these fractions for all windows (not presented here) are homogeneous, and therefore their mean is representative. We present in Figure~\ref{fig-window-users} this mean as a function of the window size.

The fraction of paedophile (IP,~port) pairs for the 2007 dataset first fluctuates for small window sizes, and quickly converges to a steady regime, very close to the overall fraction of paedophile (IP,~port) pairs in the dataset.  Notice that it is possible that a same (IP,~port) pair corresponds to several users (family computers, for instance). However, the probability that this happens {\em within a short time span} is greatly reduced.  The fact that the fraction of paedophile (IP,~port) pairs in windows of limited duration is very close to the overall fraction therefore shows that it is close to the fraction of actual detected users.

This is confirmed by the fraction of paedophile IP addresses as a function of the window size.  After some initial fluctuations, this value drops to slightly less than $0.25\,\%$, then increases linearly with the window size\,\footnote{This increase is not obvious on the figure because the slope is very small.  If the $x$ axis extended to the whole 10 weeks of measurement though, the plot would reach $0.38\,\%$ which is the overall fraction of paedophile IP addresses in the dataset.}.  Considering shorter windows therefore reduces temporal pollution.  At any given time there are nonetheless several users {\em simultaneously} using the same IP address, because they are behind a NAT for instance.  They will however use different ports, which is why the fraction of paedophile (IP,~port) pairs is always lower than the fraction of paedophile IP addresses.

The plot for the fraction of paedophile IP addresses in the 2009 dataset has the same behaviour as for the 2007 dataset, but is larger than it.  This could be because the fraction of paedophile users is larger than in the 2007 dataset.  However, as the fraction of paedophile {\em queries} in both datasets are very similar, we suspect that this is because more users use the same IP address simultaneously in 2009 than in 2007.

\subsection{Sessions}
\label{subsec-sessions}

A {\em session} is a maximal set of queries from the same IP address (or (IP,~port) pair) such that two consecutive queries are not separated by more than a given delay $\delta$.  Studying sessions reduces temporal pollution, as there will probably be a gap between the queries of two users who use the same IP address successively. On the other hand, there is no {\em a priori} reason why paedophile users would make more sessions than other users, and so we consider the fraction of paedophile sessions as an approximation of the fraction of paedophile users.

\begin{figure}[!h]
\centering
\resizebox{\columnwidth}{!}{
\includegraphics[width=\columnwidth]{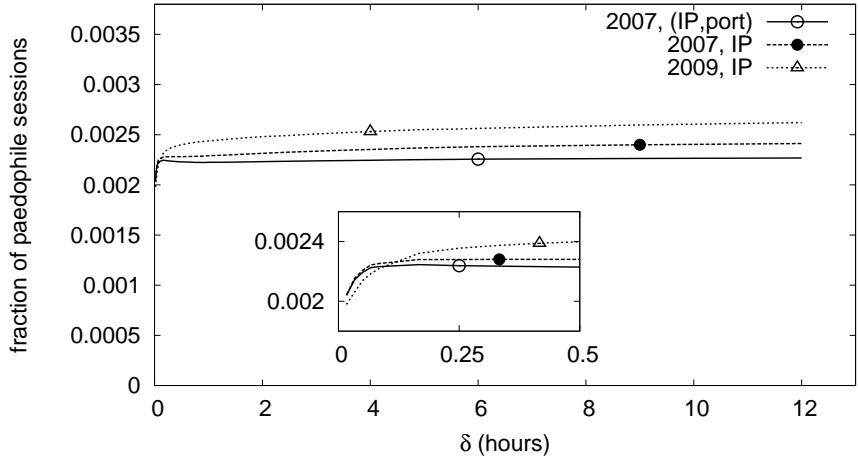}
}
\caption{
Fraction of paedophile sessions as a function of $\delta$, the maximal delay between two consecutive queries in a same session.
}
\label{fig-sessions-users}
\end{figure}

We present in Figure~\ref{fig-sessions-users} the fraction of paedophile sessions for different choices of $\delta$. The fraction of paedophile sessions for very small values of $\delta$ is not relevant, because series of queries entered by a same user then belong to several sessions. For large values of $\delta$, this fraction becomes closer and closer to the overall fraction of paedophile users in the dataset\,\footnote{If $\delta$ is equal to the measurement duration, all queries entered from the same IP address or (IP,~port) pair will belong to a single session.}.  This again confirms that considering IP addresses and connection ports seems to be enough to identify users in this dataset.

The fraction of paedophile sessions corresponding to IP addresses is higher.  This again comes from the fact that several users are simultaneously connected from the same IP address, but do not use the same port.

The fraction of paedophile sessions is larger for the 2009 dataset than for the 2007 dataset.  Again, we conjecture that this is because a higher number of users use simultaneously the same IP address.

\subsection{Inference}

The fact that the three methods used above for user quantification are in accordance shows that considering (IP,~port) pairs is relevant for identifying users in our context. The fraction of such users entering queries detected as paedophile by our tool is equal to $0.22\,\%$ in the 2007 dataset.  We now use the false positive and false negative rates of our tool to infer the actual fraction of paedophile users.

These rates give the number of queries that the tool mis-classified.  However, since we do not know which precise queries are mis-classified, we do not know what fractions of users they represent. If queries were mis-classified with uniform probability, they would correspond to a similar fraction of users. This is however probably not true, as a same user tends to enter similar queries. Therefore, if one of his/her queries is mis-classified, probably many others are.

We however establish, using the false positive rate, a lower bound for the fraction of paedophile users.  A fraction $f\primeplus{}$ of the queries detected as paedophile by the tool are in fact not paedophile, which represents a given number  $n$  of queries.  Clearly, the corresponding number of users which the tool mis-identified as paedophile is at most $n$ (it is equal to $n$ if all mis-identified queries are entered by different users).  Conversely, the tool failed to detect some paedophile queries.  If all these queries were entered by users who were nonetheless detected as paedophile (because they entered other paedophile queries which were correctly identified), then no paedophile user is missed.  The tool detected $|T\plus|=207,340$ paedophile queries in the 2007 dataset, which correspond to 112,712 different users.  The number of queries erroneously tagged as paedophile is $|T\plus|\cdot f\primeplus{} = 2,882$.  Finally, the number of paedophile users is at least $112,712-2,882 = 109,830$, which leads to a fraction of paedophile users slightly lower than $0.22\,\%$.

%Ajout Clem de derniere minute, finalement pas integre pour que le message reste clair
%Conversely, we can establish a lower bound for the fraction of paedophile users using the false negative rate.  A number $n'$ of queries are paedophile but were not detected by the tool.  The number of paedophile users which are not detected by the tool is therefore at most $n'$ (it is equal to $n$ if all queries that are not detected are entered by different users, who entered no query which was detected as paedophile by the tool).  Conversely, the tool mis-classified some none paedophile queries as paedophile.  If these queries were entered by users who also entered paedophile queries, then no non-paedophile user is erroneously detected as paedophile.  The numerical computation of this upper bound gives a fraction equal to $0.37\,\%$.

%Note however that the value we obtained for $f\moins$ (Section~\ref{sec-tool-results}) is an underestimate.  To study its impact on our upper bound, we computed the bound obtained for a value of $f\moins{}$ equal to $0.5$, which is clearly an over-estimation.  We obtain a value of $0.65\,\%$ for the upper bound.

It is not possible to establish such a lower bound for the fraction of paedophile users in the 2009 dataset, because we do not have access to the connection ports of the users. However, we observe that when we reduce the pollution caused by users successively using the same IP address (by studying measurement windows and sessions, see Figures~\ref{fig-window-users} and~\ref{fig-sessions-users}), the obtained values are close for both datasets, but larger for the 2009 dataset. We therefore estimate that a lower bound of $0.2\,\%$ of paedophile users applies to both datasets.

\section{Related work}
\label{sec-related}

Collection and analysis of large P2P traces is a very active field. Studies mainly focus on peer properties which are useful for protocol design, such as their connection time, sharing behaviour, or similarity regarding search\-ed files and geographical location, see for instance \cite{gummadi02measurement,handurukande06peer,nguyen07analysis,sen04analyzing}. Some works also analyse queries entered by users \cite{gish07geographical,klemm04characterizing} but consider limited statistics (typically query length, number, interarrival time or redundancy). Only very few studies examine user interests in detail \cite{hughes06is,shavitt09song,steel09child}. Besides, classification of search queries is an active field, with articles such as \cite{jansen} in which authors aim at identifying the proportion of sexually-related web queries.

On the other hand, many papers discuss the amount and features of paedophile activity in P2P systems but they rely on very small datasets collected manually (typically by entering a few queries and examining obtained results), \eg\ \cite{fagundes09fighting,koontz03file,waters07child}. They aim at establishing the alarming presence of paedophile activity in P2P systems, not at quantifying it, and as such cannot be compared to our work.

Up to our knowledge, only two  papers deal with the quantification of paedophile activity in a P2P system in a similar way as the work presented here \cite{hughes06is,steel09child}. Both analyse \gnutella\ traces containing keyword-based queries.

In \cite{hughes06is} the authors consider three sets of $10,000$ queries captured during three consecutive sundays in 2005. Two reviewers (whose qualifications are unknown) manually classified these queries as related to either {\em illegal pornographic content} or not (they do not focus specifically on paedophile queries, but also include incest, rape and bestiality). They conclude that $1.6\,\%$ of the observed queries are related to illegal pornography. However their dataset  is very small and therefore contains a very low number of paedophile queries, thus limiting the significance of these statistics. Moreover, their methodology for query classification is not automated nor fully specified, and it relies on two reviewers only.

Further work of this group on the same dataset proposes techniques for automatic discovery of paedophile keywords \cite{hughes08supporting}. However, as already explained, applying such methods in our context requires a significant set of known paedophile queries, which was not available before our work. Using such methods to improve our results is one of our main perspectives.

In \cite{steel09child} the author considers a set of $235,513$ queries, which is approximately $1,000$ times smaller than our datasets (again, given the small rate of paedophile queries in such datasets, this leads to estimates with limited statistical significance). He classifies queries as paedophile or not based only on the fact that they contain a keyword in a specific list, similar to our {\em explicit} list introduced in Section~\ref{sec-data}. This is not sufficient, as many paedophile queries contain no such keyword, the paedophile nature arising from a combination of non-explicit keywords. Moreover, the author provides neither his dataset nor his keyword list, thus making it impossible to reproduce his results. He concludes that almost $1\,\%$ of queries in his dataset are related to child pornography.

Finally, these contributions may be seen as pioneering but limited work on paedophile query quantification when compared to our own work. Moreover, although they discuss other interesting issues like age indications in queries and filenames, presence of sub-communities, and geographic location of users, none of them address the key question of paedophile {\em user} quantification (Section~\ref{sec-users}).

\section{Conclusion and perspectives}
\label{sec-conclu}

We addressed the problem of rigorously and precisely quantifying paedophile activity in a large P2P system. We first set up a methodology and designed a tool for automatic detection of paedophile queries. Thanks to the involvement of 21 independent highly-qualified experts of the field, we estimated its false positive and false negative rates. We collected two different datasets containing hundreds of millions keyword-based queries entered in the \edonkey\ system, and established that approximately $0.25\,\%$ of them are paedophile. We then designed several complementary methods for quantifying involved users; we established that at least $0.2\,\%$ of observed users sent paedophile queries in our 2007 dataset, similarly to our 2009 dataset.

It is the first time that quantitative information on paedophile activity in a large P2P system is obtained at this level of precision, reliability, and at such a scale. This significantly improves awareness on this topic, with important implications for child protection, policy making and internet regulation.

Moreover, our contributions open several promising directions for future work:
\begin{itemize}

\item {\em Extending our results to other systems.}
Although our observations are consistent for two datasets collected at different times and conditions, which gives them a high level of generality, it is important to obtain similar quantifications for other datasets and/or P2P systems. Indeed, different amounts of paedophile activity may occur in different systems, for various reasons (in particular, secrecy is easier in some systems than others). One may for instance collect \gnutella\ queries like in \cite{hughes06is,steel09child} and inspect them with our tool.
% Its false positive and false negative rates would have to be reevaluated for these new datasets, but this may be done with much less experts by confronting their answers to the ones presented here.

\item {\em Improving knowledge and fight against paedophile activity.}
We open the way to many studies and actions critical for understanding and
fighting paedocriminality. For instance, the low false positive rate of our
paedophile query detection tool (which may be reduced further if more false
negatives are allowed) makes it suitable for filtering at server level and more
generally on any search engine. The large sets of paedophile queries we provide
also open the way for the study of key questions regarding paedophile activity.
One may analyse for instance age indications in these queries, other topics
paedophile users are interested in, and how they start and develop their
interest in paedophile content. A deeper investigation of the combinations of
keywords used in paedophile queries may also significantly help to improve our
paedophile query detection tool. All these issues are crucial for fighting
paedocriminality and designing appropriate clinical responses.

\item {\em Applying our contributions to other contexts.} Finally, many of our contributions are not specific to paedophile activity and/or P2P systems, and could be used for other purposes. First, our methodology is directly applicable to any specific interest present in P2P systems (and more generally search engine logs) which represents a small fraction of the overall activity. This includes most deviant behaviours, such as zoophilia, rape or incest. Analysing further our datasets with appropriate methodology may also shed light on activity regarding various topics in P2P systems, like software, movie, music and pornographic contents. Notice that these datasets also allow trace-based simulations at an unprecedented scale, with great potential impact for protocol design and testing.

\end{itemize}
%\IEEEPARstart{T}{his} demo file is intended to serve as a ``starter file''
%for IEEE journal papers produced under \LaTeX\ using
%IEEEtran.cls version 1.7 and later.
%I wish you the best of success.
%\hfill January 11, 2007

%\appendices
%\section{}

\section*{Acknowledgments}
We thank the experts who helped in assessing our work, in particular Philippe Jarlov for contacts with law-enforcement personnel. We thank Ivan Daniloff for his help in collecting data, and Marcelo Dias de Amorim, Lucas Di Cioccio, Italo Cunha, Timur Friedman, Diana Joumblatt, Am\'elie Medem Kuatse, J\'er\'emie Leguay, Renata Cruz Texeira, and John Whitbeck for their help in improving preliminary versions. This work is supported in part by the MAPAP SIP-2006-PP-221003 and ANR MAPE projects. 

\bibliographystyle{plain}
\bibliography{pedo,article_sigcomm,bibliography}

%\end{IEEEbiography}
\label{theend}

\end{document}